\documentclass[a4paper, amsfonts, amssymb, amsmath, reprint, showkeys, nofootinbib, twoside]{revtex4-1}
\usepackage[english]{babel}
\usepackage{color,soul}
\usepackage[utf8]{inputenc}
\usepackage[colorinlistoftodos, color=green!40, prependcaption]{todonotes}
\usepackage{amsthm}
\usepackage{mathtools}
\usepackage{physics}
\usepackage{xcolor}
\usepackage{graphicx}
\usepackage[left=23mm,right=13mm,top=35mm,columnsep=15pt]{geometry} 
\usepackage{adjustbox}
\usepackage{placeins}
\usepackage[T1]{fontenc}
\usepackage{lipsum}
\usepackage{csquotes}
\usepackage{amsmath}
\usepackage[pdftex, pdftitle={Article}, pdfauthor={Author}]{hyperref} 
\begin{document}
\title{High-precision measurements of electric-dipole-transition amplitudes in excited states of $^{208}$Pb using Faraday rotation spectroscopy}

\author{John H. Lacy}
    \email[Correspondence email address: ]{jhl4@williams.edu}
    \affiliation{Department of Physics, Williams College, Williamstown, MA 01267, USA.}
\author{Abby C. Kinney} 
    \altaffiliation{Department of Physics, University of Chicago, Chicago, Illinois 60637, USA.}
    \author{P. K. Majumder} 
    \affiliation{Department of Physics, Williams College, Williamstown, MA 01267, USA.}

\date{\today} 

\begin{abstract}
We have completed measurements of two low-lying excited-state electric-dipole (E1) transition amplitudes in lead. Our measured reduced matrix elements of the $(6s^2 6p^2)^3P_1 \to (6s^2 6p7s)^3P_0$ transition at 368.3 nm and the 405.8 nm $(6s^2 6p^2)^3P_2 \to (6s^2 6p7s)^3P_1$ transition are 1.90(1) a.u. and 3.01(2) a.u. respectively, both measured to sub-1\% precision and both in excellent agreement with the latest \emph{ab initio} lead wavefunction calculations.  These measurements were completed by comparing the low-field Faraday optical rotation spectra of each E1 transition in turn with that of the ground state $^{3}P_0 - ^{3}P_1$ M1 transition under identical experimental conditions. Our spectroscopy technique involves polarization modulation and lock-in detection yielding microradian-level optical rotation resolution. At temperatures where direct absorption was significant for both E1 and M1 transitions, we also extracted matrix element values from a direct optical absorption depth comparison. As part of this work we designed an interaction region within our furnace which allows precise measurement of our quartz vapor cell sample temperature and thus accurate determination of the Boltzmann thermal population of the low-lying excited states that were studied.
\end{abstract}

\maketitle
\section{Introduction} \label{sec:introduction}  
Atoms and associated precision spectroscopic techniques have long been an excellent means to provide important, low-energy tests of fundamental physics associated with the standard model, as well as tests of potential new (``BSM'') physics beyond it \cite{DeMille2017, Safronova2018}.  Heavy (high-$Z$, often multi-valence) atomic species in particular, due to their enhanced sensitivity to many of these phenomena, have played an important role in experimental searches. This high-precision experimental work has developed alongside improved \emph{ab initio} atomic theory to link experimental results to interpretations and limits of the fundamental physics phenomena being targeted.  Recent atomic theory work particularly targeted at multi-valence atomic systems, and broadly applicable to many such systems, has made notable progress in accuracy in recent years \cite{Safronova2009, Safronova2013, Safronova2025}.  Unlike, for example, the alkali atom systems, far fewer general high-precision atomic structure measurements have been completed in heavy, multi-valence atoms with which to benchmark ongoing theory work and guide further refinement in these systems.
\\
\indent Building on from previous work in trivalent systems\cite{SafronovaMajumder2013,Ranjit2013,Augenbraun2016,Vilas2018}, our collaboration with the theory group of Safronova {\it{et al.}} has been recently extended to the four-electron lead (Pb) system \cite{Porsev2016,Maser2019}. Lead, along with thallium and cesium \cite{Meekhof1993,Phipp1996,Vetter1995,Wood1997}, were the subject of experimental work in the 1990's which all measured atomic parity nonconservation (PNC) at the 1\% level of precision or better. The single-valence Cs system yields a combination of experimental and similarly precise atomic theory that has long provided an important low-energy test of fundamental electroweak theory \cite{Wood1997,Porsev2009,Dzuba2012}. Recently, substantial progress in multi-valence theory work (including lead) has reduced the uncertainties in these systems several fold \cite{Porsev2016}, so that even modest further improvements in theory accuracy will have significant impacts on the quality of future tests of fundamental physics in this arena. 
\\
\indent Another important class of symmetry-violation tests involving atoms and molecules has focused on searches for potential T-violating permanent electric dipole moments (EDMs) of the electron \cite{ChuppEDM,Andreev2018,JILAEDM23}.  Lead has played an important role as a partner in diatomic molecule-based searches using PbO and PbF \cite{DeMille2000, ShaferRay2006,Baklanov2010}. Very recently, it has been proposed \cite{Stuntz2024} that a diatomic molecule including lead as one partner (paired with a gold or silver atom) has excellent potential for a future cold-molecule-based EDM search.  A first step to experimentally realizing such a cold diatomic system would be to produce laser-cooled samples of lead atoms.  In parallel with our lead vapor cell spectroscopy work, we intend to demonstrate, for the first time, transverse laser cooling in a lead atomic beam, the same apparatus which we will explore Stark-induced amplitudes and atomic polarizability measurements for additional high-precision tests of atomic theory. We are thus motivated for many reasons to pursue and expand our atomic structure measurements in this heavy, multi-valence system.
\\
\indent In 2019, we measured, for the first time, the matrix element of the very weak electric quadrupole (E2) transition within the ground-state manifold of lead and found agreement with the {\it{ ab initio}} theory at the 1\% level of accuracy \cite{Maser2019}. In the current manuscript, we describe two new measurements of lead E1 transition amplitudes, both of which with a fractional accuracy under 1\%. These measurements offer an improvement in precision of nearly an order of magnitude over previous experimental work in the lead system \cite{AlonsoMedina2001}, and show excellent agreement with latest values from our theory partners \cite{Maser2019,SafronovaPC}.
\\ 
\indent In this manuscript, having outlined the atomic structure details in Sec. II, we describe the key apparatus features and summarize data collection and analysis in Secs III.  Sec. IV discusses various potential systematic error contributions, and we conclude with a discussion of the results with a comparison to previous work and the latest theory as well as noting ongoing and future work.

\section{Atomic line shapes and \\ matrix elements} \label{sec:atomic line shapes}
 \begin{figure}
    \centering
    \includegraphics[width=.75\linewidth,trim={5.cm 0 7cm 0}]{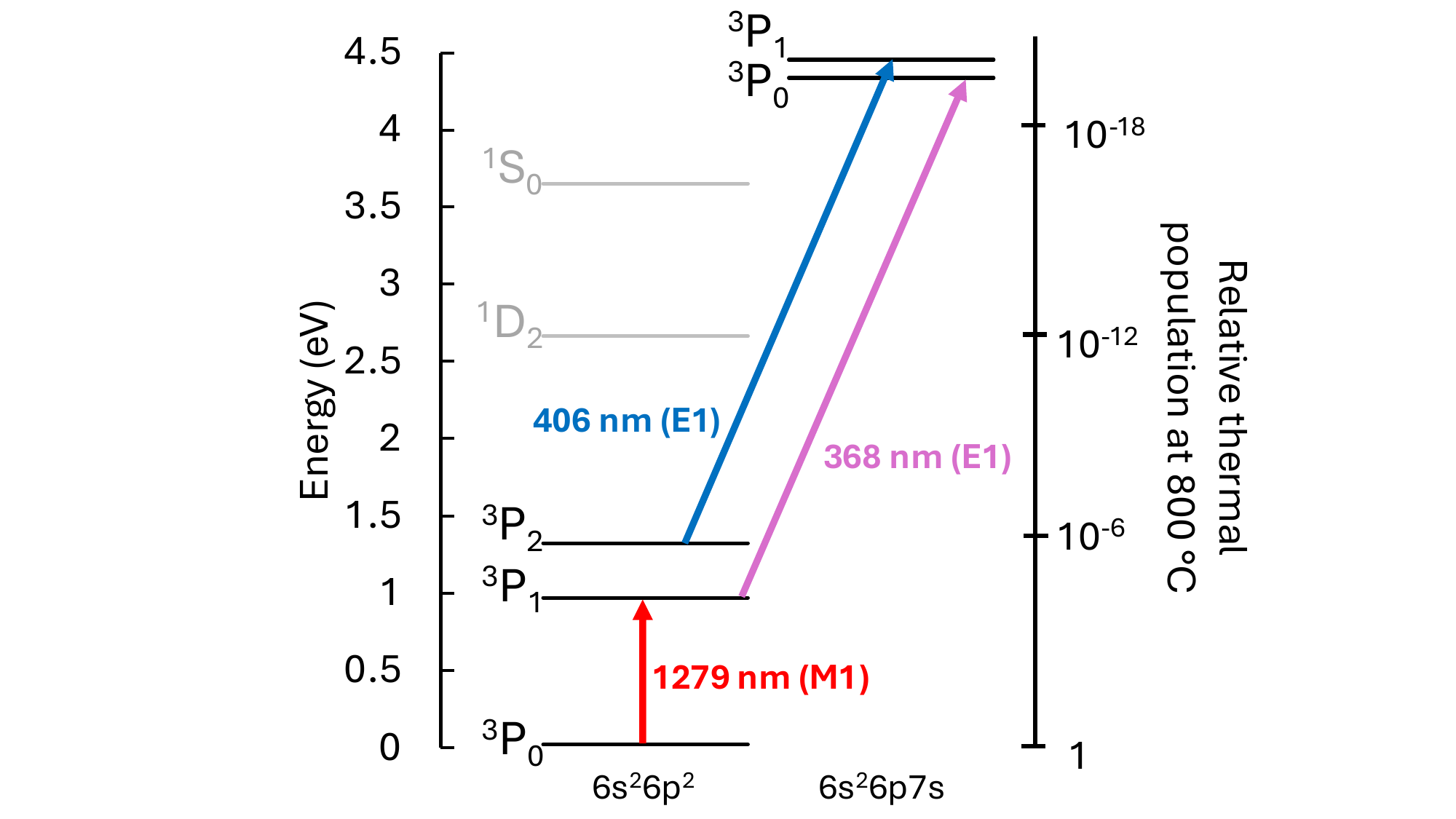}
    \caption{Low-lying energy levels and transitions in $^{208}$Pb showing relative thermal populations at 800 $^{\circ}$C.}
    \label{fig:enter-label}
\end{figure}

The relevant energy levels and transitions are shown in Fig. 1. In this study, spectral features of the 368 nm and 406 nm E1 transitions are analyzed and compared with the 1279 nm M1 transition. While the E1 transitions are intrinsically much stronger than the forbidden M1 transition, at typical experimental temperatures (700$-$900$ \, ^{\circ}$C), the spectral features turn out to be quite comparable in size owing to the much smaller thermal populations of the $^{3}P_{1}$ and $^{3}P_{2}$ levels compared with $^{3}P_{0}$ (see Fig. 1). 
\\
\indent The majority of the experimental work was conducted by Faraday rotation spectroscopy, where our polarimetry system can detect milliradian optical rotations with microradian accuracy. At high sample temperatures, where there was well-resolved direct absorption, we used analysis of transmission spectra to complement the Faraday rotation analysis. Both are described here. We note that the M1 transition amplitude does not depend on detailed wave function information, and therefore is known very precisely \cite{Porsev2016} even for the case of this multi-valence atom.  Indeed, using similar theoretical methods, M1 amplitude calculations by this theory group in other multi-valence systems has been shown to agree with experimental values at the 0.1\% level of precision or better \cite{Brewer2019, Bothwell2025}.  As such, the lead 1279 nm M1 transition offers an ideal normalization tool in our experimental determination of these E1 amplitudes.

\subsection{Transmission analysis}
\indent Starting with transmission spectroscopy, as the laser frequency, $\omega$, is scanned about an atomic transition, the measured transmitted laser intensity is given by
\begin{equation}
I(\omega)=I_{0} e^{-\alpha_0 \mathcal{P}(\omega)},    
\end{equation}
where $I_{0}$ is the incident laser intensity, $\mathcal{P}(\omega)$ is a generic normalized absorption line shape, and $\alpha_0$ is the peak optical depth, proportional to the number density, interaction length, transition frequency, and transition line strength (square or sum of squares of relevant matrix element(s)). In practice we use a (not normalized) Voigt convolution profile $\mathcal{V}(\omega)$ to describe the atomic absorption line shape which we define as
\begin{equation}
\mathcal{V}(\omega)=\frac{1}{\sqrt{2 \pi} \sigma}\int^{\infty}_{-\infty}\frac{\Gamma/2}{(\omega-\omega')^{2}+(\Gamma/2)^{2}} e^{-\frac{(\omega-\omega')^{2}}{2 \sigma^{2}}} {\mathrm{d}}\omega',   \end{equation}
where the homogeneous (Lorentzian) and Doppler spectral widths are denoted by $\Gamma$ and $\sigma$ respectively.  The resonant peak value of Eq. 2 will be a quantity that depends on the component line widths, and we refer to this as $\mathcal{V}(0)$. To be consistent with our definitions in Eq. 1, we then incorporate this amplitude factor into our definition of the peak optical depth. If we now consider the ratio of peak optical depths for our two transitions, various common factors cancel, including the (identical) path length through the vapor, and we can express this ratio as 
\begin{equation}
\frac{\alpha_{E1}}{\alpha_{M1}}=\frac{\mathcal{V}_{E1}(0)}{\mathcal{V}_{M1}(0)}\frac{N_{E1}}{N_{M1}}\frac{\omega_{E1}}{\omega_{M1}}\frac{ \langle E1 \rangle^{2}}{ \langle M1 \rangle^{2}}. 
\end{equation}
While the atomic density is the same for both observed spectra, the population ratio is not, and is given by the Boltzmann thermal population for each sub-level of the particular excited state: $N_{E1}/N_{M1}=e^{-{\Delta E}/k_{B}T}$, where $\Delta E$ is the relevant energy difference (for the case of the two relevant excited states in our current work, $\Delta E$ is 0.9695 eV for the ${^3}P_1$ state and 1.3205 eV for the ${^3}P_2$ state). From this expression, it is immediately clear that an accurate transition amplitude measurement will depend on precise cell temperature determination. 
\\
\indent The matrix element expressions suggested in Eq. 3 must be made explicit for each transition that we study, and, further, need to be connected to reduced matrix elements in atomic units to be compared to theoretical predictions.  In our polarimetry apparatus, atoms interact with linearly polarized light in all cases.  This interaction is best viewed in a spherical basis anticipating the application of magnetic fields and Zeeman-shifted resonances. Appendix A makes this connection explicit for each of the three transitions we study, employing the Wigner-Eckart theorem and a related sum rule. For the simpler 368 nm E1 transition, as well as for the 1279 nm M1 transition where there is only one transition for a given circular polarization component, it is easy to show (see Appendix A) that the ratio of reduced matrix elements is equal to the ratio of the $\Delta m=1$ matrix elements.  Thus we can find an expression that allows us to  extract the E1 reduced matrix element from various experimentally measured and fitted parameters, as well as known constants:
\begin{multline}
\langle || E1 ||\rangle = \langle || M1 ||\rangle \sqrt{\frac{\alpha_{E1}}{\alpha_{M1}}\frac{\omega_{M1}}{\omega_{E1}}\frac{\mathcal{V}_{M1}(0)}{\mathcal{V}_{E1}(0)}} \, \\ \times e^{\Delta E/2k_{B}T}\left(\frac{\mu_{B}/c}{ea_{0}} \right).
\end{multline}
We have explicitly included the physical constants that put the matrix elements into atomic units: $e$ is the elementary charge, $a_{0}$ is the Bohr radius, $\mu_{B}$ is the Bohr magneton, and $c$ is the speed of light. The exponential factor and the final units conversion factor in Eq. 4 nearly (and fortuitously) cancel, so that the observed optical depths, as well as the Faraday rotation amplitudes discussed next, are quite comparable between the M1 and E1 transitions in all of our comparison experiments.
\\
\indent For the transmission spectra of the more complicated $J=2 \to J=1$ 406 nm transition, where all the component transitions are centered at the same laser frequency (at zero magnetic field), Appendix A also shows that the sum of the absorption components stemming from each sub-level transition combine such that once again we can use Eq. 4 as written to extract the reduced E1 matrix element for this transition without the need for any additional numerical factors.  We henceforth use the shorthand $\langle || E1||\rangle_{406}$, $\langle || E1||\rangle_{368}$, and $\langle || M1||\rangle$ to refer to the three relevant reduced matrix elements.

\subsection{Faraday rotation analysis}
\indent Extracting $\langle || E1 ||\rangle$ using Faraday rotation spectroscopy requires measuring the peak optical rotation of an electronic transition (as discussed in \cite{Maser2019}). Under an applied magnetic field, $B$, the Pb sample becomes optically active and as the laser is scanned about a transition the plane of linear polarized light has a frequency-dependent ($\omega=2\pi f$) rotation according to
\begin{equation}
\Phi_{F}(\omega)=\frac{\omega l}{2 c} (n_{+}(\omega)-n_{-}(\omega)).   
\end{equation}
Here, $n_{\pm}$ refers to the real part of the Pb vapor refractive index for $\Delta m=\pm 1$ transitions. Its frequency dependence (expressed generally for a population $N$ and transition strength $\langle T \rangle^{2}$) is given by
\begin{multline}
n_{\pm}(\omega) \propto \frac{N{\langle T \rangle}^{2}}{\sqrt{2 \pi} \sigma} \int_{-\infty}^{\infty} \frac{\omega\pm \Delta \omega-\omega'}{(\omega \pm \Delta \omega -\omega')^{2}+\frac{\Gamma^{2}}{4}} \\ \times  \exp\left({-\frac{\omega'^{2}}{2 \sigma^{2}}}\right)d \omega'.
\end{multline}
The Zeeman splitting is given by $\Delta \omega=\frac{\mu_{B}g_{J}B}{\hbar}$ where $g_J$ is the appropriate Land\'{e} $g$-factor. Considering first the 1279 nm M1 and 368 nm E1 transitions, there is only one relevant $g$-factor, that of the $(6p^2) ^{3}P_{1}$ level.  For this pair of transitions, each with a single $\sigma^{+}$ and $\sigma^{-}$ component, the full, symmetric Faraday rotation line shape becomes
\begin{widetext}
\begin{equation}
\mathcal{L}(\omega) \equiv \frac{\mathcal{C}}{\sqrt{2 \pi} \sigma}\int_{-\infty}^{\infty} \left( \frac{\omega + \Delta \omega-\omega'}{(\omega + \Delta \omega -\omega')^{2}+\frac{\Gamma^{2}}{4}} -\frac{\omega - \Delta \omega-\omega'}{(\omega - \Delta \omega -\omega')^{2}+\frac{\Gamma^{2}}{4}} \right) \times  \exp\left({-\frac{\omega'^{2}}{2 \sigma^{2}}}\right)d \omega'.
\end{equation}
\end{widetext}
In practice, we would insert the appropriate $g$-factor (for the $(6p^2) ^{3}P_{1}$ level $g=1.501(2)$ \cite{Wood1968,Porsev2016}) as well as the magnetic field value for a given run, extracting the homogeneous width from the Faraday line shape fit, and determining the fitted amplitude factor $\mathcal{C}$, which is then implicitly a function of the component widths, the $g$-factor, and the magnetic field. We next employ the same angular momentum arguments just discussed to produce our ratio of reduced matrix elements. Our final expression for the E1 reduced matrix element as extracted from the Faraday analysis can thus be expressed in terms of $\mathcal{C}_{E1}/\mathcal{C}_{M1}$, the M1 reduced matrix element, the transition frequency ratio, and the Boltzmann factor as follows:
\begin{equation}
\langle \norm{E1} \rangle =\sqrt{\frac{\mathcal{C}_{E1}}{\mathcal{C}_{M1}}\frac{\omega_{M1}}{\omega_{E1}}} \,e^{\Delta E/2k_{B}T}\left( \frac{\mu_{B}/c}{e a_{0}}\right)\langle \norm{M1} \rangle.    
\end{equation}

\begin{figure}
    \centering
    \includegraphics[width=1\linewidth]{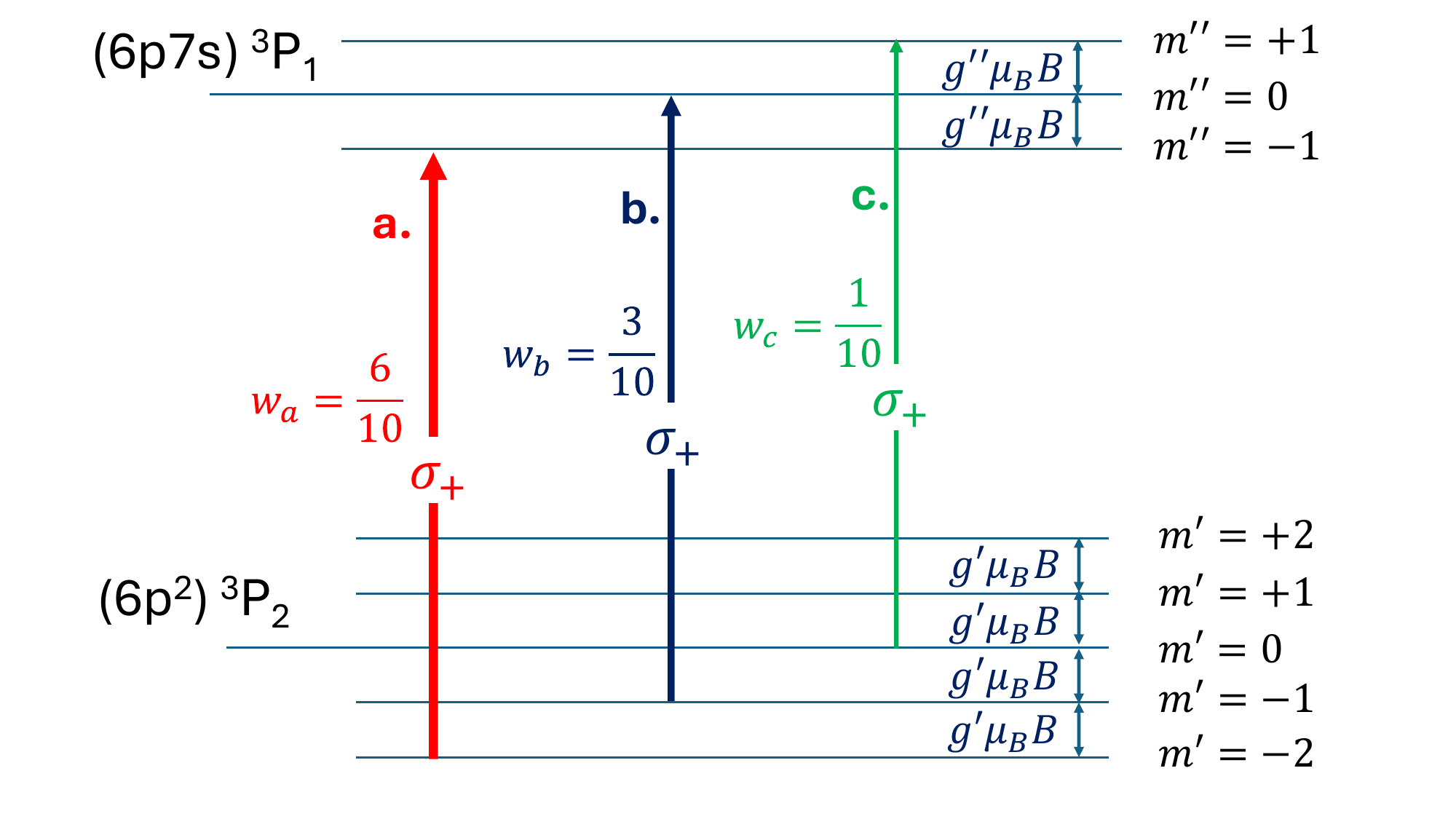}
    \caption{The three $\sigma_{+}$ contributions to the 406 nm spectral feature showing appropriate weightings and Zeeman splittings. Experimental \cite{Wood1968, Lurio1970} and theoretical \cite{Porsev2016} values for $g'$ and $g''$ can be found in the literature. For clarity, the three corresponding $\sigma_{-}$ transitions are not shown.}
    \label{fig:enter-label}
\end{figure}

\indent We note that for these small applied magnetic fields, the Zeeman shifts of the magnetic sublevels are negligible compared to the overall energy difference $\Delta E$ so that we assign the identical Boltzmann population factor to each. 

\indent We finally turn to the more complicated 406 nm transition Faraday line shape.  Referring again to the discussion in Appendix A as well as Fig. 2, this ($J_{i}=2\rightarrow J_{f}=1$) transition comprises a trio of $\sigma_{+}$ and corresponding $\sigma_{-}$ components each weighted according to the relevant Clebsch-Gordan coefficient and each Zeeman-shifted by different amounts according to the two different $g$-factors in play (labeled $g'$ for the lower state and $g''$ for the upper state in Fig. 2). Thus we must expand our two-dispersion fitting function (Eq. 7) to include six terms with individual weighting factors and $g$-factors. Referring to the labels a, b, and c in Fig. 2, these weighting factors (which sum to unity) are: $w_a = 6/10, w_b = 3/10, w_c = 1/10$ and the corresponding `effective' $g$-factors are: $g_a = (2g' - g''),\, g_b = g',\, g_c = g''$. For analysis of the 406 nm transition Faraday data, we use this expanded, properly normalized six-dispersion curve model, and retaining the overall model amplitude factor $\mathcal{C}$ in our fitting process, we are then able to employ Eq. 8 as written to extract $\langle ||E1||\rangle_{406}$.  
\\
\indent Given that all of the Zeeman shifts here remain small compared to the overall dispersion curve linewidth, one can make an excellent two-dispersion approximate model for this transition, using the weighted average of each set of three dispersion curves and a single `weighted-average' $g$-factor.  This allows us to employ the same line shape model as for the 1279 nm and 368 nm transitions, and we find that the final results obtained using the `exact' vs `approximate' fits to the 406 nm transition Faraday data differ by less than 0.1\% in all cases.

\section{Experimental Details} \label{sec:experimental}
\subsection{Apparatus}
    \begin{figure*}
        \centering
        \includegraphics[width=1\linewidth, trim={0cm 2cm 0cm 1cm}]{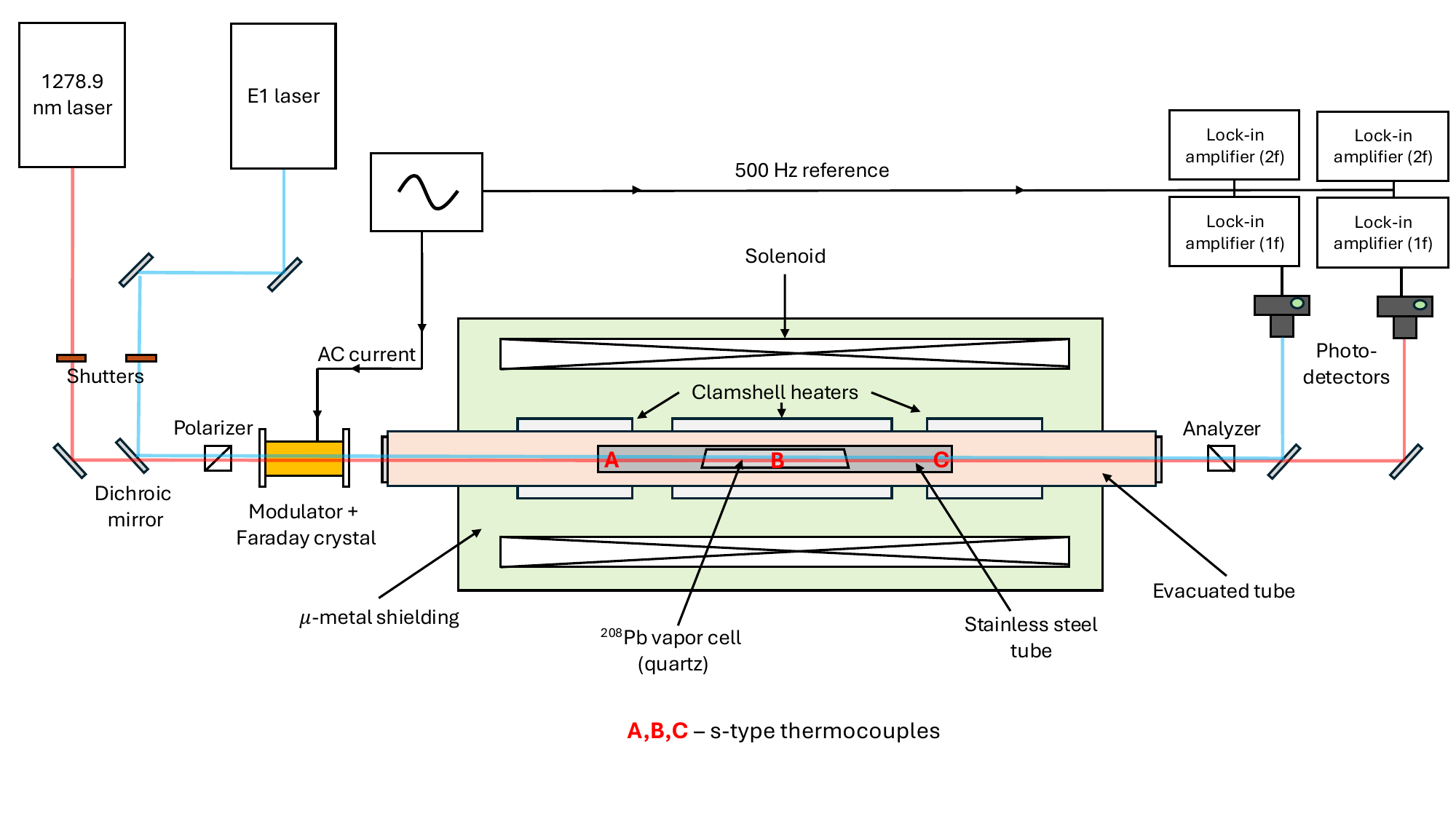}
        \caption{Cross-sectional schematic of the experimental apparatus.}
        \label{fig:enter-label}
    \end{figure*}

The basic experimental apparatus is shown in Fig. 3 and is quite similar to that used in \cite{Maser2019}. In an effort to ensure temperature uniformity, we placed a 15 cm-long quartz cell containing isotopically enriched $^{208}$Pb (purity$=99.97 \%$) in the center of a home-built vacuum furnace. The cell windows are angled to avoid the possibility of multi-passing of the laser beams due to reflections. The furnace consists of three pairs of clamshell heaters surrounding a 1 m-long ceramic tube, evacuated and back-filled with 20 Torr of argon. Each individual clamshell heater is driven by the amplified output of a function generator delivering a 10 kHz sine wave. The maximum total power output of the furnace is roughly 900 W. The large thermal mass of the furnace, in conjunction with a software \emph{p-i-d} control loop which controls the amplitude of the function generator driving the heater amplifiers, ensures that after a warm-up period of several hours, the furnace temperature remains stable to better than $1 \, ^{\circ}$C. 
\\
\indent The cell environment is monitored by 3 s-type (Pt:PtRh) precision thermocouples fixed at points A, B, and C within the stainless steel cylinder that encloses the Pb vapor cell (see Fig. 3). The thermocouple voltage is converted to a temperature reading using a Stanford Research Systems SRS 630 precision thermocouple reader, which corrects for dissimilar metal contact voltages at the instrument junctions by accurately monitoring the temperature of the connection pads. The 30 cm-long stainless cylinder and its endcaps help to provide a stable, uniform thermal environment for the vapor cell. Slight adjustments to the individual heater currents allow optimization of temperature uniformity. Thermocouple accuracy was confirmed (at least at lower temperatures) by putting them in contact with both a water/ice slurry and boiling water, which gave agreement with expected temperatures at the $0.1 \, ^{\circ}$C level. This is well within the quoted thermocouple accuracy over its full temperature range of $\pm{1.5} \, ^{\circ}$C. Thermocouple variance was tested at a range of temperatures by clamping them to a single point, but no difference in thermocouple reading was observed beyond the $\pm{0.5} \, ^{\circ}$C level at any time during the heating cycle.
\\ \indent 
Surrounding the heaters are several layers of insulation, water cooling pipes, and a solenoid delivering 22.0(2) G/A. This was accurately determined (see Appendix B) by measuring the Zeeman splitting of the $(6p^{2})^{3}P_{1}$ level whose $g$-factor is known to high precision \cite{Wood1968}. A $\mu$-metal cylinder surrounding the entire furnace limits any external transverse magnetic fields in the interaction region to less than 10 $\mu$T.
\\ 
\indent The optical and signal processing setup is also similar to that of previous work in Pb \cite{Maser2019}, and is substantially the same for both the 368 nm and 406 nm E1 transition amplitude measurements. In each case, our 1279 nm laser (Sacher Lasertechnik) drives the reference M1 transition, while the beam from a home-built E1 external cavity diode laser is aligned using a dichroic mirror to traverse the same beam path as the 1279 nm laser. Path overlap was optimized by maximizing the signal \emph{enhancement} from two-step excitation of the 1279 nm and 368 nm transition. (For the 406 nm transition, this enhancement was observed and optimized through a secondary effect, whereby the 1279 nm M1 laser enhances the $^{3}P_{1}$ level which further populates the $^{3}P_{2}$ level through collisions). We further ensure geometric overlap with the use of small collimating apertures placed 1.5 m apart through which both laser beams must pass. 
\\
\indent The lasers (which in the experiment are alternately blocked and unblocked using computer-controlled shutters) pass through the atomic sample located in between two Glan-Thompson polarizers. The emerging light beams are separated by a second dichroic mirror before being directed into their respective photodetectors. Between the first polarizer and the atomic sample is a magneto-optic element which consists of a high-Verdet-constant material.  The principal difference in our two experimental setups was the use of two different crystals: a 1 cm-long, 1 cm-diameter TGG glass was used for the 406 nm measurement, and a 5 mm-long, 5 mm-diameter CeF$_{3}$ unixial single crystal element for the 368 nm measurement. The latter was used for its favorable transmission properties in the UV \cite{Lacy2024}, though its twenty-fold smaller Verdet constant at IR wavelengths led to somewhat higher optical rotation noise levels for the M1 Faraday data here (yet still below $10 \, \mu \mathrm{Rad}/\sqrt{\mathrm{Hz}}$). 
\\ 
\indent 
In each case, the relevant Faraday modulator crystal is subjected to a sinusoidally-varying (500 Hz) magnetic field of roughly 75 Gauss that modulates the light polarization. The detected signals are then demodulated using four lock-in amplifiers (Stanford Research Instruments, model SR810 DSP) referenced to $1f=500$ Hz and $2f=1000$ Hz for each laser. As discussed in \cite{Maser2019,Lacy2024}, the $2f$ component provides a signal directly proportional to the transmission signal (with no optical rotation information), while the ratio of the $1f$ and $2f$ lock-in signals gives a signal proportional to the atomic Faraday optical rotation, $\Phi_{F}(\omega)$ when we apply a longitudinal DC magnetic field to the atoms, so that:
\begin{equation}
\frac{V_{\mathrm{lock-in}}^{1f}}{V_{\mathrm{lock-in}}^{2f}}=\kappa \Phi_{F}(\omega), 
\end{equation}
where $\kappa$ is a calibration constant determined experimentally in the course of data collection, as described below. 
\\ 
\indent Along with the $1f$ and $2f$ lock-in signals, a Fabry-P\'{e}rot transmission spectrum was recorded to provide the relative frequency axis of each collected Faraday spectrum. A 600 MHz resonant electro-optic modulator and RF synthesizer was used to determine the free spectral ranges (FSRs) of the M1 (500.7(3) MHz) and E1 (500.4(4) MHz) confocal cavities (not shown in Fig. 3). 
\\
\subsection{Data collection}
\indent
For each E1/M1 comparison experiment, we performed sequential frequency scans of the two lasers (both upward and downward going) across the respective transitions both with and without the DC Faraday magnetic field.  The eight-scan sequence required roughly one minute to complete and was repeated (in various orders) over the course of roughly 45 minutes. For each scan, we also collected Fabry-P\'{e}rot transmission spectra, and recorded the value of the applied DC solenoid current, as well as the temperature readings for each of the three thermocouples within the interaction region. 
\\
\indent In order to calibrate the measured $1f/2f$ lock-in ratio in terms of absolute radians (effectively measuring the constant $\kappa$ in Eq. 9), both before and after each such set of runs we induced a common angular `uncrossing' in our polarimeter by using a computer-controlled stepper motor and differential micrometer to effect a small, controlled polarization rotation of roughly 5 mRad in our second polarizer (analyzer). This calibration procedure was performed twice before and twice after each data set.  Since we measure the \emph{ratio} of amplitudes, the exact magnitude of the mechanical uncrossing is not important, only that it is the same for both lasers. The precision and reproducibility of this procedure was aided by the microradian-level control of our mechanical lever-based polarizer mount \cite{Lacy2024}. As discussed below, we assigned calibration errors based on both the reproducibility of repeated calibration steps and the variance of the before vs. after data-set calibration exercises. 
\\
\indent
We typically re-optimized the optical system after one data set. Over the course of several weeks, we changed a variety of experimental parameters such as the magnitude of the DC Faraday field, the E1 laser power, laser sweep conditions, etc.  Data were taken over a wide range of sample temperatures between 700-$850 \, ^{\circ}$C which corresponds in both comparison experiments to an order-of-magnitude change in Pb vapor density, and a factor of four change in the thermal Boltzmann factor noted in the previous section.  In the end, we collected and analyzed more than 500 individual spectral scan pairs for each comparison experiment taken over a wide variety of experiment conditions including a range of different DC magnetic fields and E1 laser powers.  A comprehensive analysis of both statistical and potential systematic errors associated with the overall data sets is discussed below.

\section{Data analysis and results} \label{sec:data analysis and results}
\subsection{Line shape fitting procedure}

\begin{figure*}
    \centering
    \includegraphics[width=1\linewidth,trim={0cm 4cm 0cm 8cm}]{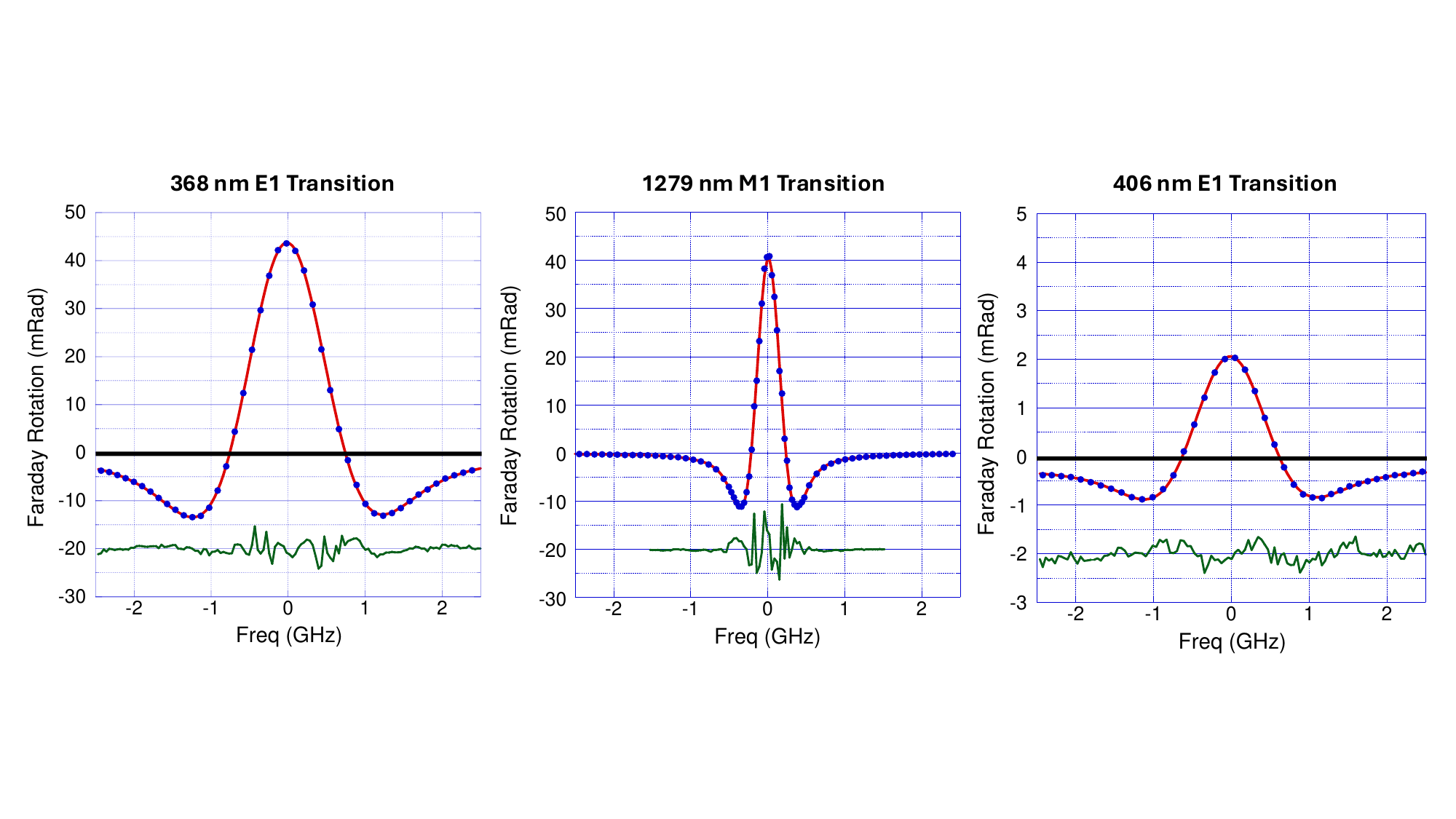}
    \caption{Sample data (blue markers) for single (field on - field off) Faraday rotation scans for all three transitions at 800$\, ^{\circ}$C and 44 Gauss. Each scan displayed required roughly 30 seconds to acquire. The line of best fit is shown in red, the expanded fit residuals are shown below each line shape (see text). While the 1279 nm and 368 nm optical rotation amplitudes are of comparable size, the amplitude of the 406 nm transition is an order of magnitude smaller due to the reduced thermal population of $(6p^{2})^{3}P_{2}$ level compared with the $(6p^{2})^{3}P_{1}$ level. }
    \label{fig:enter-label}
\end{figure*}

\begin{figure}
    \centering
    \includegraphics[width=1\linewidth, trim={5cm 3cm 5cm 2cm}]{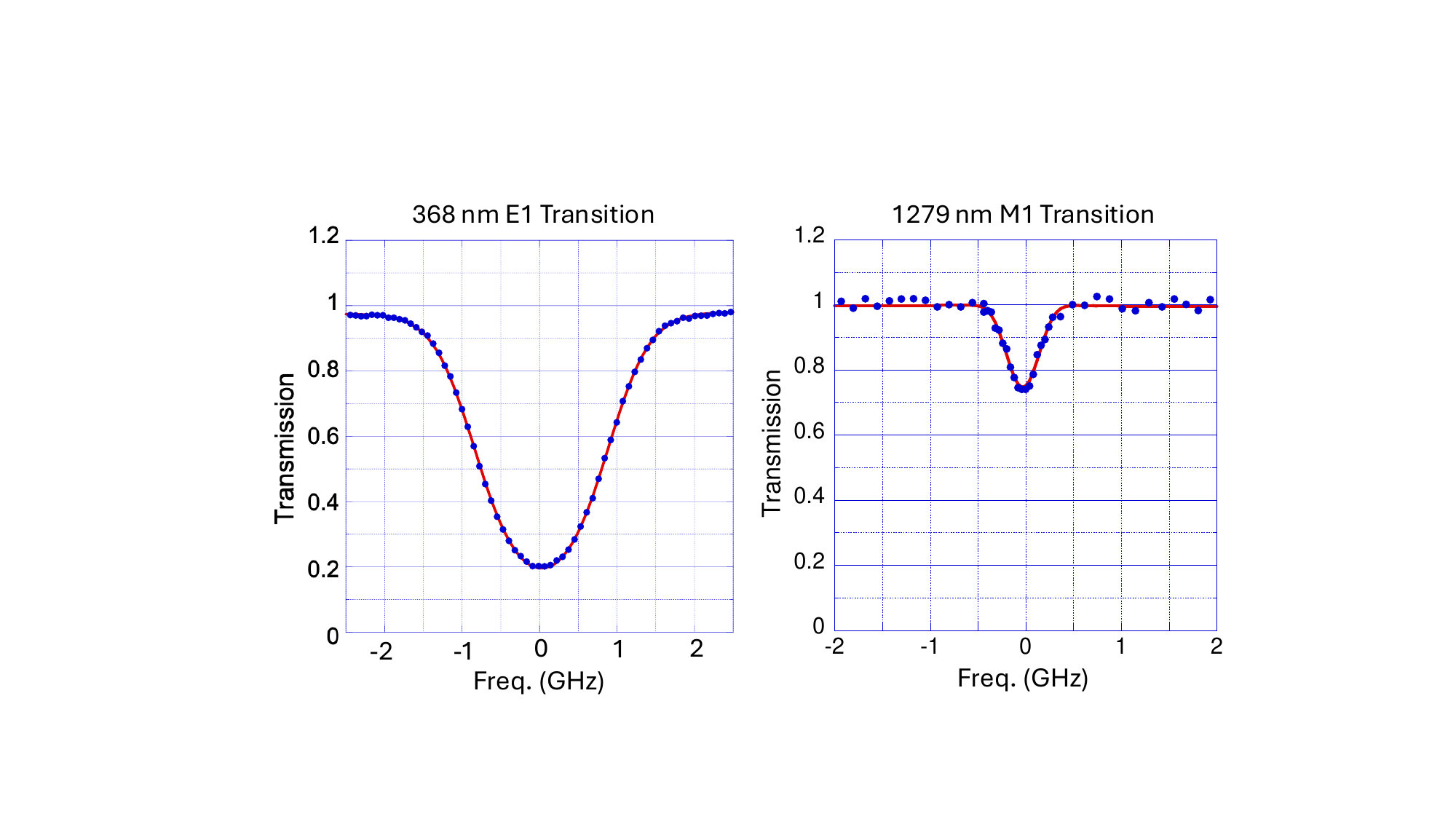}
    \caption{Sample single transmission data scans (blue markers) and line of best fit (red) at 825$\, ^{\circ}$C for the 368 nm (left) and 1279 nm (right) transition. The somewhat higher noise level for the 1279 nm transition signal is a consequence of the much smaller $2f$ modulation depth due to the reduced Verdet constant of the CeF$_{3}$ crystal at 1279 nm.}
    \label{fig:enter-label}
\end{figure}

\indent As a first step in preparing our spectra for analysis, a linearized and calibrated frequency axis of each spectra is obtained by fitting the Fabry-P\'{e}rot transmission data to an Airy distribution. Nonlinear scan behavior was accounted for by expanding the Airy argument up to a fourth-order polynomial. Higher-order polynomials were explored but were found to have no effect on the extracted E1 transition matrix element. Next, we subtract the lock-in $1f/2f$ signals with magnetic field off from the associated field-on spectra which removes frequency-dependent polarization backgrounds unrelated to the atoms. We then used the average value of the before- and after-data set calibration steps to provide background-subtracted spectra with amplitudes in units of radians. Using the calibrated frequency scale, we can separately create transmission spectra suitable for direct optical depth analysis using the $2f$ lock-in output for the case of zero magnetic field (that is, using our `background' scans).  
\\
\indent We establish a nominal temperature to associate with each scan for two reasons. First, as described next, we are able to fix the Doppler width of our spectra simplifying the fitting process. Second, as shown in Eqs. 4 and 8, we require a temperature value to evaluate the Boltzmann factor that allows us to ultimately extract the E1 amplitude values. For the large majority of the data collected, the three thermocouple readings within the vapor cell enclosure were consistent within the quoted $\pm 1.5 \, ^{\circ}$C thermocouple accuracy.  Here we assigned the central thermocouple value (labeled ``B'' in Fig. 3) to be the nominal temperature.  We ultimately assigned an overall temperature-uncertainty systematic error to our results, as described in Sec. IV.C.  
\\ 
\indent Finally, we also establish a DC magnetic field value to associate with each Faraday scan so that we can fix the Zeeman splitting in our fits as well.  In Appendix B, we describe our method to calibrate the magnetic field based on a Doppler-free, two-step spectroscopy scheme using the 1279 nm M1 laser overlapping with the 368 nm UV laser to probe the Zeeman structure of the $(6p^{2})^{3}P_{1}$ whose magnetic $g$-factor is very well known (see {\cite{Wood1968} and references therein). Sec. IV.C. further discusses the accuracy of this method and potential systematic errors associated with possible inaccuracies in assigning magnetic field values to our fits.
\\ 
\indent We use a computationally efficient series approximation using Dawson functions \cite{MeekhofThesis} further discussed in Appendix C to fit our Faraday and transmission spectra to appropriate Voigt convolutions. Referring to Eq. 7, having fixed the Doppler width based on the determined temperature and the Zeeman splitting of the dispersive curves, and with a calibrated frequency axis in place, our Faraday fitting function `floats' only the homogeneous width, the overall amplitude, and an overall DC offset from imperfect background subtraction. For the infrared M1 transition, the homogeneous width, due primarily from collisional broadening, is roughly 30 MHz, or about 10$\%$ of the Doppler width. For the E1 transition line shapes, both the homogeneous and Doppler widths are larger, and the ratio of the two is quite similar to that of the M1 line shape. Of note, for the E1/M1 comparison involving the UV laser, where we employed the uniaxial CeF$_{3}$ crystal modulator, the E1 fits took into account a possible small ellipticity in the line shape due to crystal birefringence (making one dispersive component slightly larger than the other), as discussed in \cite{Lacy2024}.  With careful alignment we are able to minimize this effect, and typical fits found ellipticity values consistent with zero. Sec IV.C discusses systematic error investigations associated with experimental parameters, calibration and details of the line shape fitting procedures.
\\
\indent When fitting direct transmission spectra, we similarly fixed the Doppler widths and then fit the (magnetic field-off) $2f$ lock-in signals to the function described by Eqs. 1 and 2.  We allowed for a frequency-dependent linear intensity background in the fits to account for incident laser power variation across the resonance.  From these fits, we again extracted homogeneous width components and also peak optical depths from which the E1 amplitudes are obtained. Importantly, we saw good agreement of the homogeneous widths as extracted from the Faraday and the transmission line shapes. 
\\
\indent 
For experimental conditions where direct transmission analysis was statistically feasible, these quite distinct analysis procedures provide a powerful test of potential hidden systematic errors. For example, the transmission fits do not rely on calibration of the signal amplitude or magnetic field, whereas the Faraday technique provides a zero-background signal whose very high intrinsic precision allows data collection over a much wider range of temperatures and vapor densities. For the 368 nm transition, originating from the $(6p^{2})^{3}P_{1}$ state, we were able to obtain high-quality transmission fits and extract accurate E1 amplitude values for the data taken at 800$\, ^{\circ}$C and 825$\, ^{\circ}$C. Because of the order-of-magnitude smaller thermal Boltzmann factor for the $(6p^{2})^{3}P_{2}$ state, we were able to obtain transmission results with precision sufficient for useful statistical comparison to the Faraday results only at the highest temperature (850$\, ^{\circ}$C) for 406 nm E1 transition.  
\\
\indent Fig. 4 shows samples of single, background-subtracted Faraday rotation scans with solid lines representing model fits.  These sample scans were taken with nominal values of magnetic field and laser power, and at a temperature in the middle of our experimental range.  Each spectrum required roughly 30 seconds to acquire. Shown below the data and fit line are residuals, expanded by a factor of 10. The increased statistical noise in these residual plots align with the slopes of the Faraday curves, and is consistent with a frequency `jitter' of 1-2 MHz, typical of our ECDL systems, as further discussed in \cite{Maser2019}. Fig. 5 shows a sample of direct transmission spectra for the 368 nm E1/M1 comparison experiment taken at 825$\, ^{\circ}$C. Data points and model fit are again shown.

\subsection{Summary of results}
\indent After determining amplitude values from the Faraday fits (\emph{i.e.} the $\mathcal{C}$ constant in Eq. 7), we insert the nominal temperature for the particular spectrum pair into the Boltzmann factor expression in Eq. 8 along with the other constants, and the accurately-known \cite{Porsev2016} M1 reduced matrix element $\langle || M1 || \rangle = 1.293(1)$ a.u. to obtain a value for the E1 reduced matrix element. When we analyze a full set of scans within one data set (typically 25 up and 25 down spectrum pairs), the fractional statistical standard error for the average extracted E1 transition amplitude is at or below $1\%$.  Similarly, from a pair of fitted transmission spectra, we insert the values of the optical depths into Eq. 4 along with the thermal Boltzmann factor to again obtain E1 transition amplitudes. 
\begin{figure}
    \centering
    \includegraphics[width=1\linewidth,trim={1cm 1cm 0cm 1cm}]{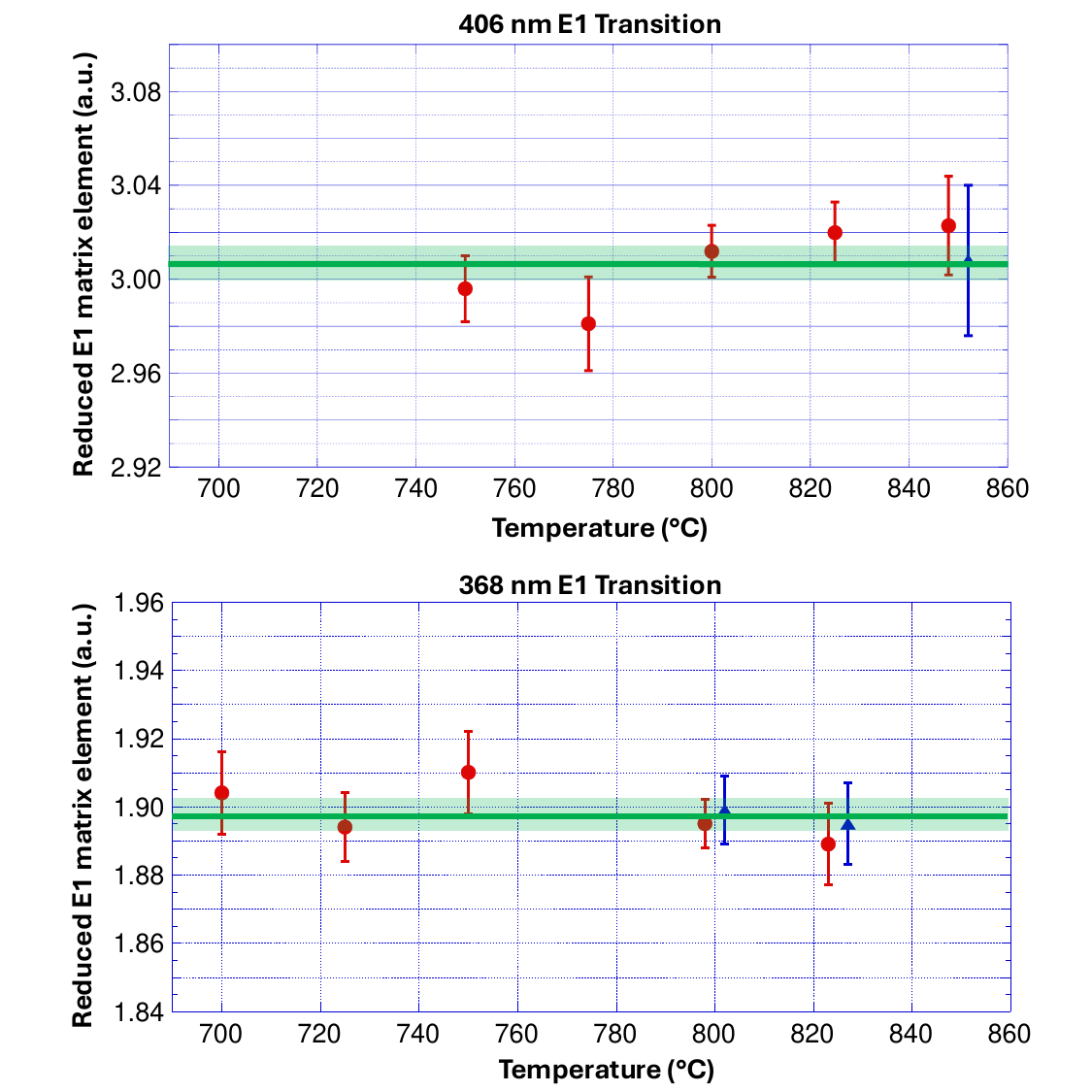}
    \caption{Measured E1 matrix elements for different temperatures. The red circular data points show results from fitting Faraday spectra, while the blue triangular data points show transmission spectroscopy measurements. The horizontal green line shows the overall E1 amplitude mean value, while the green band represents $\pm 1\sigma$ statistical uncertainties in the mean.}
    \label{fig:enter-label}
\end{figure}
\\
\indent In Fig. 6 we have assembled results of all our data sets with associated error bars grouped in order of increasing cell temperature/density for each of the E1 transitions including Faraday results (red circles) as well the transmission line shape results (blue triangles). Over an order of magnitude change in Pb vapor density, we find no resolved trend in either E1 amplitude versus temperature, and that the transmission analysis is in excellent agreement with the Faraday results where common analysis was possible. Fig. 6 includes statistical errors only. For each data set we found the average (weighted by the statistical errors derived from each fit) and then added the associated fractional uncertainty of that set in quadrature with the uncertainty in our mechanical angle calibration procedure (typically a few tenths of 1\% fractional uncertainty, and common to all scans within a set). For the latter contribution, we considered the difference between before-set and after-set calibration procedures. 
\\
\indent As indicated in Table 1, the final fractional statistical error for each E1 transition amplitude was near 0.2\%. We determined each final value by taking the weighted mean value of all data sets. As an alternative method, we measured the mean of the (unweighted) histogram consisting of every individual measurement of the transition amplitude. For both transitions, we found excellent agreement in the mean values derived from these two methods. Our final values (including statistical errors only) are: $\langle || E1 || \rangle_{368} = 1.897(4)$ a.u. and $\langle || E1 || \rangle_{406} = 3.008(7)$ a.u.

\subsection{Systematic error discussion}
\indent We took two different approaches in exploring our results for potential systematic errors. First, we split the collected data into discrete subsets taken with differing experimental conditions and looked for evidence of statistically significant dependencies on various parameters. Examples of this include considering results from upward vs. downward laser scans, results for various values of DC magnetic fields used for Faraday rotation, different E1 laser powers, and, as plotted in Fig. 6, results grouped by cell temperature. Second, to explore possible line shape model or fitting inaccuracies, we refit subsets of data with altered model parameters or fitting procedures and looked for changes in the extracted values for our E1 matrix elements. 
\\\indent Although the large majority of the complete body of data were Faraday rotation-based, we nevertheless separated the analysis of the Faraday and transmission results to account for distinct systematic errors. While the Faraday results were far more statistically precise, they were also subject to more potential systematic errors. We here enumerate and summarize the systematic error contributions to the Faraday data only in Table 1. We note that the final two entries in the table (in red text) are factors which directly scale the extracted E1 amplitude (see Eqs. 4 and 8), and would potentially bias both Faraday and transmission data in the same direction. We therefore took the approach of analyzing statistical and all relevant systematic contributions for Faraday data and transmission data separately, took the weighted average of these, and then finally combined that resultant error with the contributions due to temperature and M1 amplitude uncertainty in quadrature to obtain the final quoted uncertainty discussed in Sec. V. Below we focus largely on the Faraday data systematics investigations in particular.

\begin{table}
\begin{ruledtabular}
\begin{tabular}{lcr}
\textrm{Source} &
\textrm{$\Delta \langle \, E1 \, \rangle_{368}$}& \textrm{$\Delta \langle \, E1 \, \rangle_{406}$} \\
\colrule
Statistical error (Faraday only) & 0.22\% & 0.23\% \\
\hline
{\it{Line shape / fitting}} &  & \\
Frequency linearization & 0.04  & 0.06 \\
Free spectral range & 0.03  & 0.05 \\
Linear background inclusion & 0.13 & 0.14 \\
Include/discount wings & 0.30 & 0.15 \\
Ellipticity & 0.15 & $-$ \\
2 vs 6 dispersive curves & $-$ & 0.03 \\
\hline
{\it{Calibration}} &  &  \\
Pre/post angle calibration & 0.15 & 0.16 \\
Magnetic field uncertainty & 0.05 & 0.05 \\
$g$-factor uncertainty & $-$ & 0.18 \\
\hline
\textcolor{red}{Boltzmann factor/temperature} & 0.46 & 0.58\\
\textcolor{red}{$\langle \norm{M1} \rangle$ uncertainty} &  0.08 & 0.08 \\
\end{tabular}
\end{ruledtabular}
\caption{\label{fig:epsart} Error budget for both E1 transition amplitude measurements as derived from the Faraday analysis. A similar error analysis approach was applied to the transmission results (see text). The final two entries (red text) are separated to indicate that these potential systematics would affect all results in a `common mode' fashion and so are applied after averaging results from transmission fits and Faraday fits. All uncertainties are expressed in percentages.}
\end{table}

\subsubsection{Line shape model and fits}
As we have done in all recent spectroscopy work \cite{Maser2019,Vilas2018} we explored potential line shape related systematics by refitting subsets of data with higher-order polynomials to account for laser scan non-linearity, and exploring the effect of small calibration errors due to Fabry-P\'{e}rot FSR uncertanties (see Table 1). 
\\ 
\indent Line shape fits were also performed both with and without the off-resonant `wings' of the Faraday spectra, and while the observed changes were small, the effect on the 368 nm Faraday spectra was slightly larger, due to the complications associated with small ellipticity and asymmetry caused by the CeF$_{3}$ crystal modulator \cite{Lacy2024}. For the 368 nm Faraday line shape fits, this ellipticity $-$ which arises from small differences in the amplitudes of the $\sigma_{+}$ and $\sigma_{-}$ contributions to the (otherwise symmetric) Faraday spectra $-$ was explicitly accounted for by the introduction of a small ellipticity parameter in the model. Typically, these effects were statistically unresolved, which was achieved by ensuring the lasers were aligned along the optic-c axis of the crystal (see \cite{Lacy2024} for further details). To explore possible correlations of this ellipticity with Faraday amplitude and eventually matrix element determination, we compared fits of a representative batch at 800$\, ^{\circ}$C both with and without an ellipticity fit parameter and found changes at the 0.1\% level. Accounting for such ellipticity was not required for the 406/1279 nm $\langle\norm{E1}\rangle$ comparison, since the biaxial TGG crystal modulator used there exhibits no birefringenece. 
\\
\indent Another potential feature in all of our Faraday line shapes is due to possible imperfect background subtraction which we explored by re-fitting subsets of data with a linear background slope parameter. The results of these studies are included in Table 1. We note that the first three entries in the table apply also to the analysis of possible systematic errors in our transmission spectra, and we took a similar approach to re-fitting these spectra, with error contributions at a similar 0.1\% level or below.
\\
\subsubsection{Calibration uncertainties}

\indent As discussed in Sec. III, the $1f/2f$ lock-in signal was calibrated in units of milliradians via a computer control of a precision differential micrometer which in turn produced very small polarizer rotations via a lever mechanism \cite{Lacy2024}. Calibration was performed before and after each collected dataset. In estimating potential systematic effects here, we focused on the consistency of the calibration \emph{ratio} by studying the typical variance between the before-data set vs. after-data set calibrations taken 30-60 minutes apart.
\\ \indent In our Faraday rotation line shape analysis, we enter the particular Zeeman shifts as fixed, `known' parameters while we extract the line shape amplitude and homogeneous widths from the fits themselves. If either the magnetic field calibration or the values of the relevant $g$-factors are incorrect, this could lead to systematic errors in key fitted parameters. The various $g$-factors discussed and referred to in Sec II have been measured \cite{Wood1968,Lurio1970} and are in good agreement with theory values \cite{Porsev2016}. We use the following consensus values: $g = 1.501(2); \, g'=1.276(2); \, g''=1.352(3)$. Residual errors in these values do not contribute to the 368/1279 nm comparison since they involve a common $g$-factor. To study the impact of these uncertainties on the 406/1279 nm comparison, using both experimental and simulated data, we refit the spectra with $g$-factors altered by an amount consistent with the error bars stated above, and assessed the impact of these uncertainties on the eventual computed value for the E1 amplitude (see Table 1). 
\\
\indent Regarding magnetic field calibration, our \emph{in situ} method for field calibration at the 1\% level of accuracy is described in Appendix B.  We can similarly explore the consequence of incorrectly fixed magnetic field values through refitting experimental or simulated data with various slightly altered fixed values for the field. In the limit of the small-field approximation, the effect of a changing magnetic field would exactly cancel in our ratio measurement; here there is some differential effect since the Zeeman splitting represents a different fraction of the overall line shape width for the M1 vs E1 cases.
\\ \indent The most significant uncertainty in our entire experimental project, equally relevant for Faraday and Transmission data analysis, is associated with temperature calibration and the effect that this has on the thermal Boltzmann factor appearing in Eqs. 4 and 8. As discussed in Sec. III, the sample temperature was monitored with three s-type thermocouples located in the stainless stell enclosure containing our vapor. These have a quoted accuracy of $\Delta T= \pm$ 1.5$\, ^{\circ}$C. In most cases, the three thermocouple readings were in agreement within this intrinsic error. In these instances, we take the temperature uncertainty as $\Delta T/\sqrt{3} \approx \pm 0.9$$\, ^{\circ}$C. For the 368 nm and 406 nm transitions, the corresponding uncertainty in the relevant Boltzmann factors leads to an $\langle \norm{E1}\rangle$ uncertainty of 0.46\% and 0.58\% respectively at 800$\, ^{\circ}$C where the bulk of the data were collected (this percentage error depends very weakly on temperature over our full experimental range). In the few cases where there was a small but resolvable temperature \emph{gradient} across the atomic sample, we assigned an additional temperature uncertainty (up to $\pm$ 1$\, ^{\circ}$C) whose additional propagated uncertainty in the transition amplitude value was added in quadrature with fit errors and calibration errors, effectively reducing the statistical weight of that run. This additional gradient uncertainty was estimated by finding the difference between the central thermocouple reading, $T_{B}$, and the integrated average of the temperature distribution across the length of the cell. The temperature distribution was found by fitting the three thermocouple readings, $T_{A,B,C}$, considering their locations with respect to the cell (see Fig. 3), and assuming a smooth, continuous function of position.
\begin{figure}
    \centering
    \includegraphics[width=\linewidth,trim={0cm 0cm 0cm 0cm}]{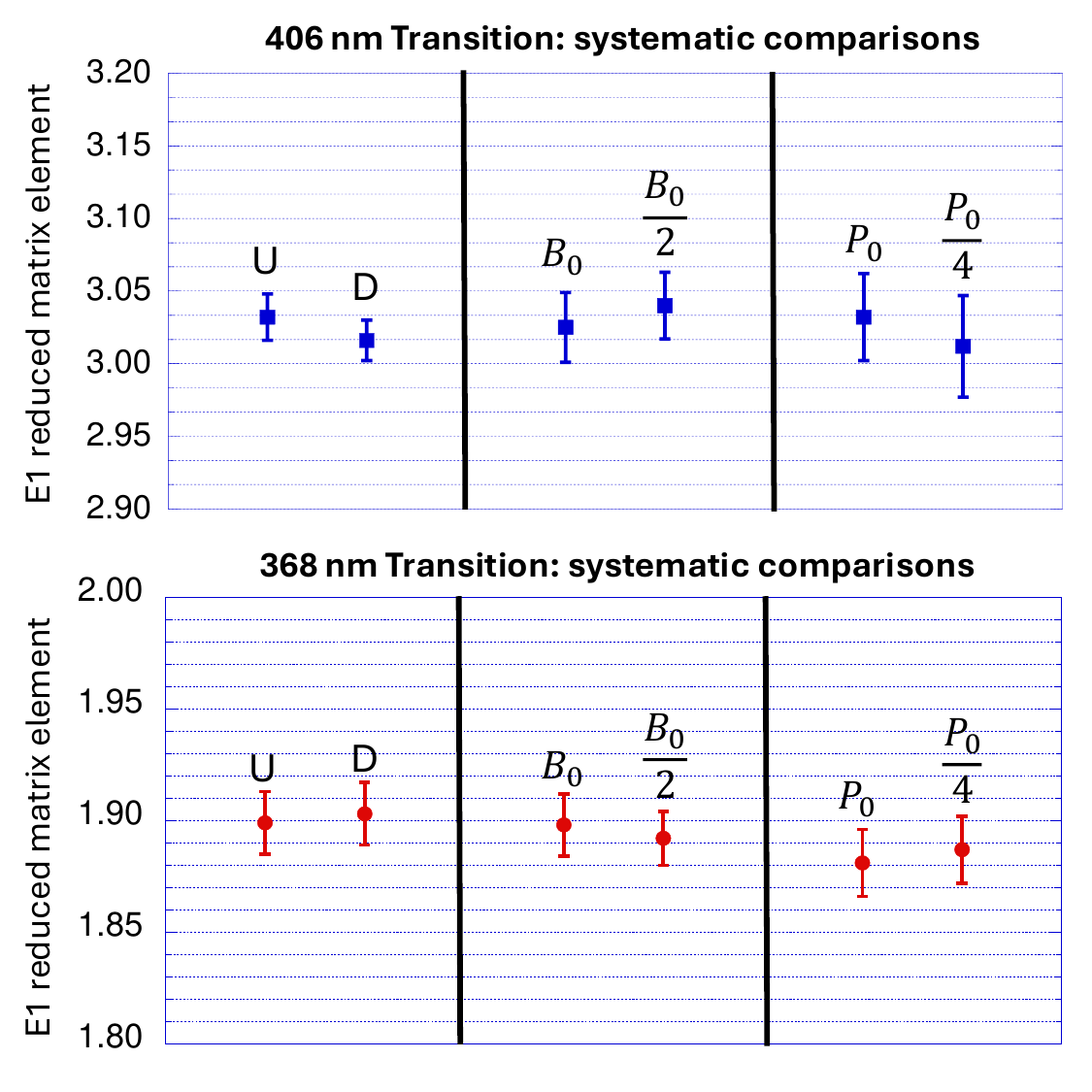}
    \caption{Response of extracted $\langle \norm{E1} \rangle$ values to changes in several experimental parameters. Values are derived for Faraday spectroscopy for the case of each comparison experiment. The left hand panel shows comparisons of laser upscans (U) vs. downscans (D). The middle panel shows $\langle \norm{E1} \rangle$ values for two magnetic fields (where $B_{0}=44$ G). The right hand panel explores two significantly different E1 laser powers ($P_{0}\approx0.5$ mW). In all cases, we see no statistically resolved changes.}
    \label{fig:enter-label}
\end{figure}

\subsubsection{Experimental conditions}
We explored potential dependence of extracted $\langle\norm{E1}\rangle$ values on several key experimental parameters including cell temperature, magnetic field, laser power, and laser scan direction by repeating data sets with a single parameter altered, holding all others fixed. A sample of comparisons of data subsets for both transition ratio experiments is shown in Fig. 7.
\\
\indent As noted above and shown in Fig. 6, we see no statistically resolved trend in our extracted E1 amplitude values over a wide range of temperature values. We extended our temperature range upward to be able to simultaneously collect and analyze direct transmission data which could contribute statistically meaningful comparative results. We also note that the temperature directly determines the Doppler width in our line shape analysis, which we enter as a fixed parameter in the Voigt convolutions, and exploring this broad range of temperatures allowed us to confirm that this procedure produced consistent results.\\
\indent
Though we studied the impact of magnetic field calibration errors on our results through refitting of data, we also explored this question experimentally, varying the applied field over the range of 20-60 Gauss and studied the E1 amplitude results as function of these changes, and we show in Fig. 7 the result of consecutive data sets taken with all other experimental conditions held constant and a `halving' of the DC field, which shows no statistically significant change in the extracted E1 value. 
\\ \indent Variation of E1 laser power could lead to systematic errors through optical saturation effects or possible line shape changes due to power broadening.  The intensity of both the UV and blue lasers at the location of our vapor cell are quite low $-$ our nominal intensities were below $1$ mW/cm$^2$, a factor of roughly 50 below the saturation intensity for these transitions. We nevertheless explored the effects of a substantial change in laser power (here a factor of four reduction), and observed no resolved change in our E1 amplitude results.  Such saturation effects are not a concern for our 1279 nm laser given the much weaker intrinsic line strength of this magnetic dipole transition.
\\ \indent Finally, to verify direction-dependent scan consistency of our laser systems, as well as the accuracy of our Fabry-P\'{e}rot-based scan-linearization procedure, we explicitly explored potential $\langle\norm{E1}\rangle$ variation as a function of scan direction and again found no resolved difference between upscans and downscans for either experiment.  This last potential systematic error source was also explored for our direct transmission data analysis with the same conclusion.

\section{Discussion} \label{sec:discussion}

After including statistical and systematic errors, and combining our Faraday and transmission results as described in the previous section, we arrive at the following final experimental values for the two Pb matrix elements studied here $-$ for the $(6p^{2}) ^{3}P_{1} \to (6p7s) ^{3}P_{0}$ 368 nm transition we find: $\langle || E1 || \rangle_{368}=1.897(11)$ a.u., and for the $(6p^{2}) ^{3}P_{2} \to (6p7s) ^{3}P_{1}$  406 nm transiton we find: $\langle ||E1|| \rangle_{406}=3.008(22)$ a.u.. From Table 1 it is clear that the temperature uncertainty of our vapor cell sample dominates the final combined error in each case. As it turns out, for any given temperature error, we find a three-fold smaller fractional uncertainty contribution if we consider the \emph{ratio} of our E1 matrix element measurements (there is partial, but not complete, cancellation of Boltzmann factors and associated uncertainties). However, this significant reduction in the overall error budget is counterbalanced by the increased error resulting from taking the ratio of our two otherwise uncorrelated measurements. In the end, we find that 
\begin{equation}
\frac{\langle ||E1|| \rangle_{406}}{\langle ||E1||\rangle_{368}} = 1.586(9),
\end{equation}
and we see that the fractional error in \emph{this} quantity is roughly the same as that for our individual matrix element determinations.
Notably, the reduced dependence on cell temperature in the ratio value offers a useful cross-check on our temperature determination procedure generally. A significant systematic temperature error beyond our quoted temperature uncertainties would have biased both E1 amplitude measurements in a common-mode fashion, but would not be apparent in the ratio value where the temperature-related error contribution is small. The fact that both the individual measurements and the ratio are in very good agreement with the theory predictions reinforces our confidence in the temperature determination and quoted uncertainty.
\\ \indent We are now in a position to compare our new measurements to previous experimental work as well as the latest theoretical work. Substantial progress has been made in recent years in \emph{ab initio} modeling of electron wavefunctions in multi-valence atomic systems. Current, state-of-the-art `CI+all order' computational techniques have yielded both improved precision, and excellent agreement with recent benchmark experiments in three-valence indium and four-valence lead from our group \cite{Vilas2018,Maser2019,Safronova2025}. We here quote the following recommended \emph{ab initio} theory values for our two matrix elements. For the 368 nm transition: $\langle || E1 || \rangle_{th}=1.92(6)$ a.u. \cite{Maser2019} and for the 406 nm transition: $\langle || E1 || \rangle_{th}=2.99(9)$ a.u. \cite{SafronovaPC}. The estimated 3\%  theory uncertainties arise from comparison of results using an alternative CI-based approach, as well as estimates of the size of several small uncalculated corrections \cite{SafronovaPC}. Our results now establish an important new experimental sub-1\%-precision benchmark against which to compare future theory work.  We note that in principle, given parallel theoretical treatments of these two matrix element calculations, a theoretical \emph{ratio} of these quantities (which, in analogy to Eq.10, would produce the ratio: 1.56) will exhibit a somewhat reduced fractional error since both the lower and upper levels of these transitions are associated with common fine structure manifolds.
\begin{figure}
    \centering
    \includegraphics[width=\linewidth]{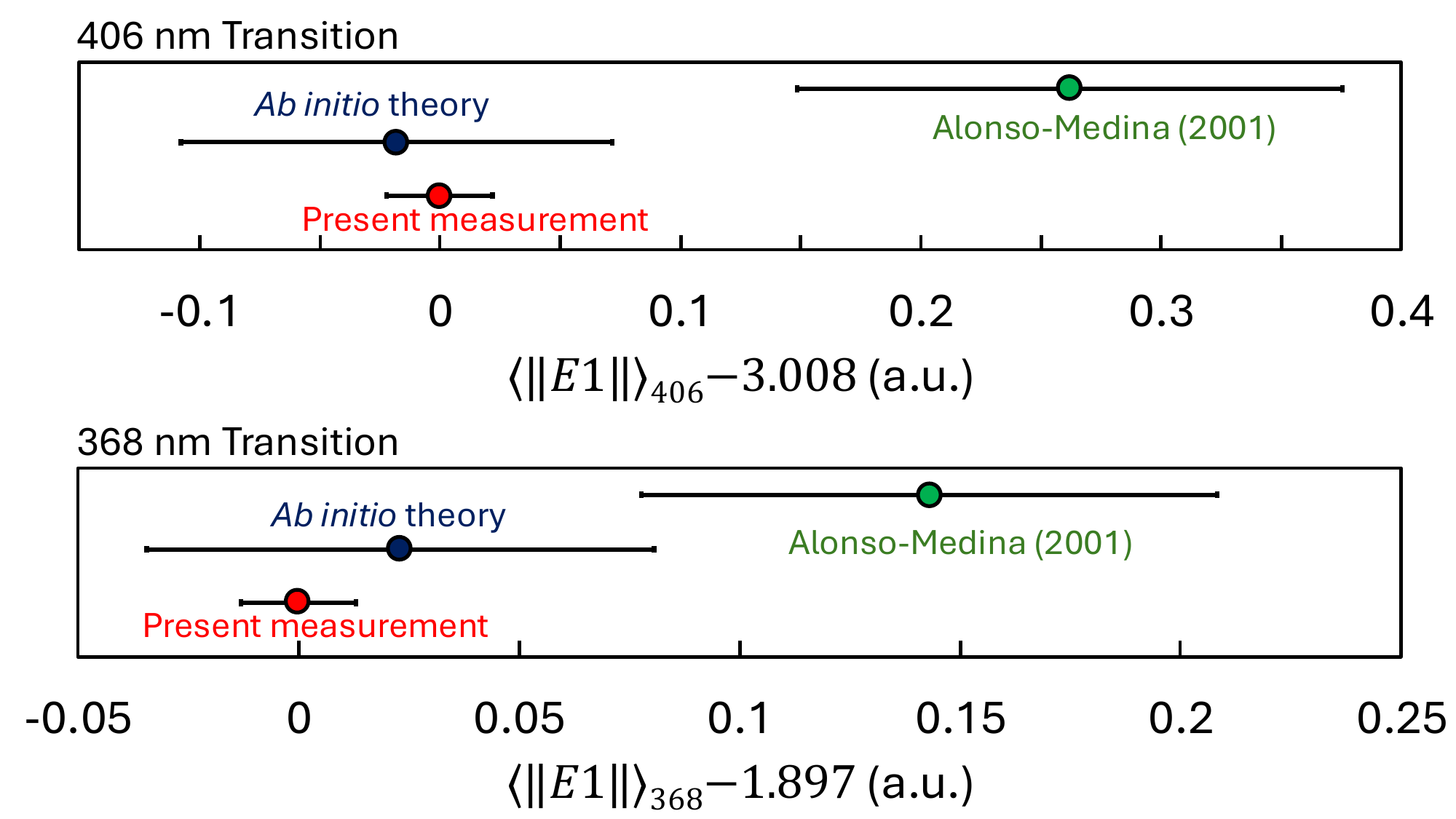}
    \caption{Final measurements of $\langle  \norm{E1} \rangle$ values (red) showing excellent agreement with the latest theory work from Porsev and Safronova \cite{Maser2019,SafronovaPC} (blue). Older experimental values from Alonso-Medina \cite{AlonsoMedina2001} are also shown in green.    \label{fig:enter-label}}
\end{figure}
\\
\indent Fig. 8 shows our results for the two matrix elements along with the theory predictions, as well as the most precise previous experimental results.  The older experiment \cite{AlonsoMedina2001} studied lifetimes of excited states and related emission line strength ratios in laser-induced lead plasmas. Older, less precise experimental determinations of $\langle\norm{E1}\rangle$ are not included in Fig. 8, but details of that work can be found in \cite{Penkin1963, Salomann1966, Lotrian1979, AlonsoMedina1996}. Our results show an increase in precision by a factor of nearly an order of magnitude compared with previous experimental work, and both of our measurements lie below the older values by roughly 2$\sigma$, where $\sigma$ is the combined standard deviation of the old and new measurements. The fact that the older experimental values, which used identical experimental approaches to each other, lie systematically above both our new measurements and the latest theory predictions suggests the possibility of a common experimental systematic error in that earlier work.  With experimental accuracy now exceeding that of theory, our measurements provide a benchmark for future theory work in this four-valence system.

\section{Concluding Remarks} \label{sec:conclusions}
We have completed measurements of the $(6p^{2})^{3}P_{1}\to (6p7s)^{3}P_{0}$ and $(6p^{2})^{3}P_{2}\to (6p7s)^{3}P_{1}$ excited-state transitions in $^{208}$Pb offering significant improvement in precision compared with previous experimental measurements. Our measured results show excellent agreement with the {\it{ab initio}} theory, and our ratio of E1 values provides an experimental benchmark for theory with sub-1\% precision. Our experimental method of measuring excited-state transition amplitudes using high-precision Faraday and transmission spectroscopy of E1 and M1 transitions opens up the possibility for future measurements in other heavy element species. As a notable example, the low-lying level structure of thallium, which includes an accurately known ground-state infrared M1 transition, points to a similar scheme of E1/M1 comparison through Faraday rotation measurements of multiple low-lying thallium excited states with the expectation of similar sub-1\% precision.
\\ \indent Currently we are also pursuing additional experimental benchmarks through measurements of lead polarizabilities using transverse atomic beam spectroscopy in a similar fashion to our earlier work in thallium and indium \cite{Ranjit2013,Augenbraun2016,Vilas2018,Doret2002}.  In the course of this atomic-beam-based work, we intend to pursue transverse laser cooling of a Pb atomic beam making use of the (optical cycling) 368 nm $(6p^2)^3P_1 \to (6p7s) ^{3}P_0$ transition that has been studied in the present work. The laser cooling of lead would be a first step towards the assembly of cold Pb-based diatomic molecules for potential next-generation EDM studies.

\section*{Acknowledgements} \label{sec:acknowledgements}
    We gratefully acknowledge support of the National Science Foundation through the RUI program and grant no. 1912369. We thank C. Yang, R. Blakey, and G. E. Patenotte for their experimental contributions to various aspects of this project, and thank J. Mativi and M. Taylor for technical support in design and construction of several pieces of apparatus for the experiment. We thank M. S. Safronova for valuable discussions, and B. L. Augenbraun for helpful comments on the manuscript.

\appendix*
\section{A: Reduced matrix elements}

In all cases, our lasers are linearly polarized perpendicular to the quantization axis defined by the propagation direction and the DC Faraday magnetic field.  Thus we can parametrize our transitions as due to equal parts $\sigma_{\pm}$ circular polarization components. We label our matrix elements as $|n,J,m \rangle$, where $n$ refers to the atomic state, $J$ to the total electronic angular momentum, and $m$ to its projection.  We note that where there are multiple initial states involved we assume that (such as for the $(6s^2)^3P_1$ and $(6p^2)^3P_2$ excited states relevant for the 368 nm and 406 nm transitions respectively) all sub-levels are equally populated according to the relevant thermal Boltzmann factor. 
\\ \indent First we consider the relevant linestrengths for our three transitions, relating the measured matrix elements to the reduced matrix elements we seek.  For a general spherical tensor operator $T^{k}_{q}$ of rank $k$ and projection $q$ the Wigner-Eckart theorem states
\begin{multline} 
\ \left|\left\langle n_f,J_{f,}m_f\middle|\ T_q^k\middle|\ n_i,J_{i,}m_i\right\rangle\right|^2= \\
 \left|\left\langle J_{f,}m_f\middle|\ k,q;J_{i,}m_i\right\rangle\right|^2 \frac{|\langle n_f,J_f||T^k||n_i,J_i\rangle|^2}{ 2J_f+1},
\end{multline}
where the first term on the RHS is the square of the relevant Clebsch-Gordan (CG) coefficient, and the numerator of the second term is the $m$-independent reduced matrix element.  We consider three dipole transitions in our experiment, so that $k=1$ in all cases. Taking $q=1$ ($\sigma_+$ transition) for specificity, the relation between the measured matrix element and the reduced matrix elements for the $J_{i}=0 \to J_{f}=1$ M1 transition, and for the $J_{i}=1 \to J_{f}=0$ 368 nm E1 transition using Eq. A.1 become
\begin{multline} \notag
|\langle 6p^2,1,1|M1|6p^2,0,0|\rangle|^2 = \\
\frac{1}{3} |\langle 6p^2,1|| M1|| 6p^2,0\rangle|^2,
\end{multline}
and
\begin{multline} \notag
|\langle 6p7s,0,0|E1|6p^2,1,-1\rangle|^2 = \\
\frac{1}{3} |\langle 6p7s,0||E1\||6p^2,1\rangle|^2 .
\end{multline}
While in one case the factor of 1/3 arises from the $(2J_f+1)$ factor, and in the other from the square of the CG coefficient, it is clear that the ratio of the reduced matrix elements is identical to the ratio of the matrix elements we measure experimentally.  For this pair of transitions, there is exactly one $\Delta m = +1$ component in each case, which makes clear why we can employ the reduced matrix elements as written in Eqs. 4 and 8 of Sec. II.
\\ \indent For the 406 nm E1 transition involving a $J_{i}=2 \to J_{f}=1$ transition there are three transitions that contribute to the absorptivity for the case of the transmission spectra and to the peak amplitude for the Faraday spectra. First consider the magnetic field-free transmission case. Again considering the $\Delta m=+1$ transitions, though the sublevels are equally populated, each transition is weighted by different CG factors. For the transmission spectrum analysis, where for zero field all transition strengths simply add, we can appeal to a convenient sum rule which states,
\begin{multline}
\sum_{{m_i},{m_f}} \ \left|\left\langle n_f,J_{f,}m_f\middle|\ T_q^k\middle|\ n_i,J_{i,}m_i\right\rangle\right|^2 = \\
\frac{|\langle n_f,J_f||T^k||n_i,J_i\rangle|^2}{ 2k+1},
\end{multline}
where again $k=1$ for this dipole transition.  Since the sum over all possible component transitions is exactly what we observe experimentally, and since Eq. A.2 clearly yields the same factor of 1/3 on the RHS of this relationship as above, we can again use Eq. 4 as written for ratio of $\langle ||E1||\rangle_{406}$ to $\langle ||M1||\rangle$ for this more complicated dipole transition as well.
\\ \indent For the 406 nm transition Faraday analysis, each transition needs to be individually treated since the spectral Zeeman shifts are unique to each, hence the six-dispersion model discussed in Sec II of the manuscript.  However, Eq. A.1 can be used to `weight' each term by the square of the appropriate CG coefficient, which are 6/10, 3/10, and 1/10 for the $\sigma_+$ transitions originating from the $(6p^2) ^3P_2, m=-2,-1, 0$ sublevels respectively.  With these properly weighted trios of dispersion curves associated with the $\sigma_+$ and $\sigma_-$ polarizations, we can expand Eq. 7 in the text to include six terms, retaining the overall amplitude fitting factor $\mathcal{C}$ as was done for the other two transitions, as discussed in the text.
\appendix*
\section{B: Magnetic field measurement}

The magnetic field was measured {\it{in situ}} by observing the Zeeman splitting of the $(6p^{2}) ^{3}P_{1}$ level. With both the 1279 nm and 368 nm lasers overlapping and passing through the heated vapor cell, the 1279 nm laser (locked to the field-free peak of the first-step transition) was ``chopped'' at 2 MHz using an acousto-optic modulator. The transmission signal of the scanning 368 nm laser was then demodulated by an RF lock-in amplifier. This modulation frequency is sufficiently fast to avoid velocity-changing collisions, yielding a single Doppler-free peak (with Lorentz width of around 100 MHz) at zero field. 
\\
\indent As the magnetic field is increased, this single peak splits into two (see Fig. 9). The observed frequency splitting is a combination of the $\sigma_{\pm}$ transitions of both the 1279 nm and 368 nm transitions, and so there is a `Doppler enhancement' of the observed frequency splitting related to the ratio of laser wavelengths. This can be understood as follows. With 1279 nm laser tuned to the center of the M1 resonance line, consider first the lower-frequency peak of the 368 nm laser lock-in spectrum originating from the $^3P_1 (m=+1)$ state. Here the 1279 nm laser first excites an (off-resonant) $\sigma_{+}$ transition. The velocity-class of atoms that are excited are moving towards the laser with velocity $v=\delta f_{z} \cdot \lambda_{1279}$, where $\delta f_{z}$ is the Zeeman shift of the $m=1$ level. Due to selection rules, the overlapping 368 nm laser can only excite this same group of atoms with a $\sigma_{-}$ transition. Since these atoms are moving towards the UV laser, the atoms ``see''  blue-shifted light, and so the resonance occurs at a frequency shifted below resonance by
\begin{equation} \tag{B.1}
\delta f_{368}^{\sigma_{+}}=-\delta f_{z}-\frac{v}{\lambda_{368}}=-\delta f_{z}\left(1+\frac{\lambda_{1279}}{\lambda_{368}} \right).    
\end{equation} 

\begin{figure}
    \centering
    \includegraphics[width=1\linewidth,trim={0.cm 2cm 0cm 2cm}]{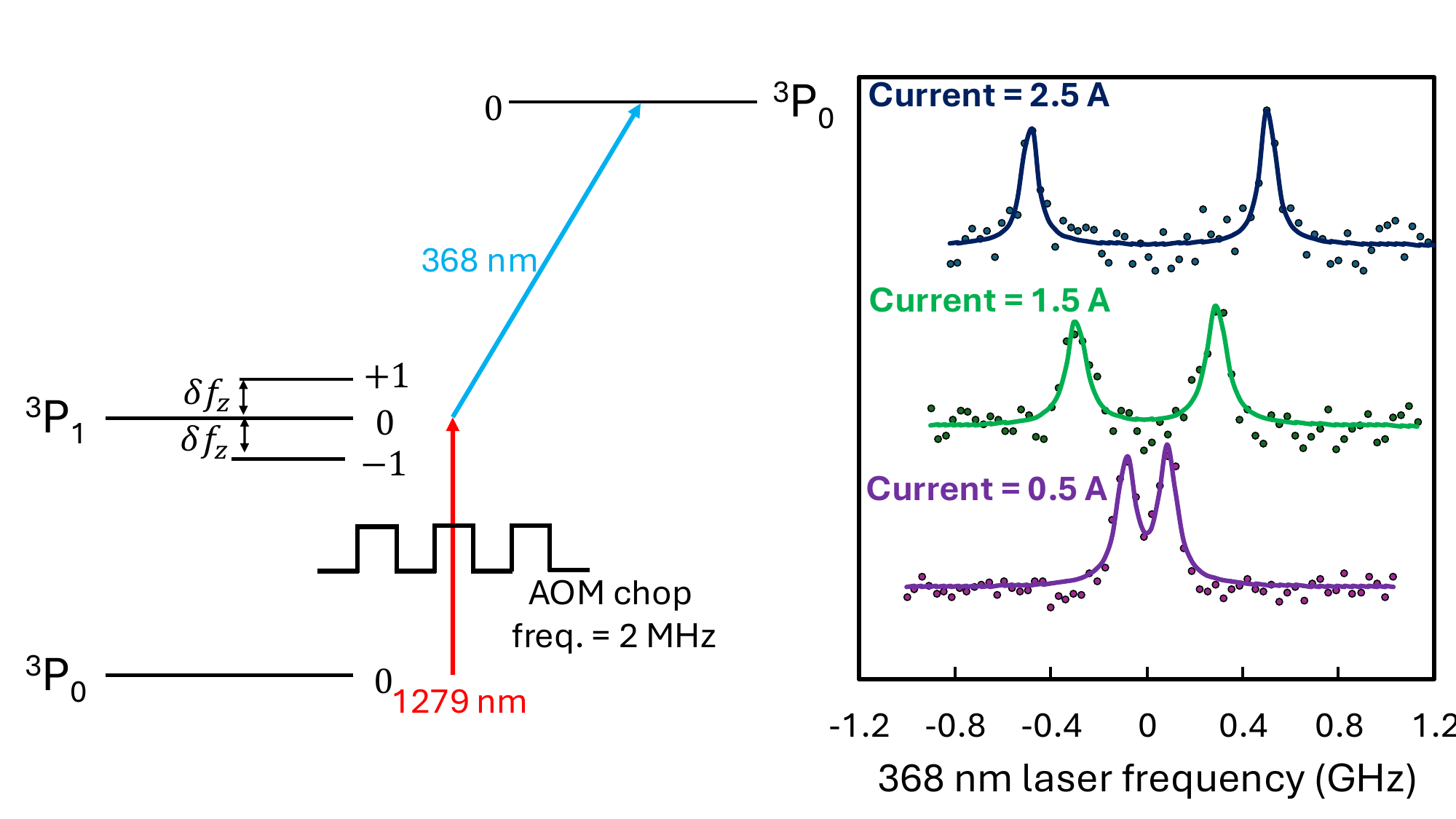}
    \caption{A schematic of the {\it{in situ}} magnetic field measurement (see text). The insert show sequential scans of the 368 nm (second-step) laser for several different solenoid currents.}
    \label{fig:enter-label}
\end{figure}

\indent By symmetry, the second lock-in peak corresponds to a $\sigma_{-}$ M1 transition followed by a $\sigma_{+}$ E1 transition that is shifted by $\delta f_{368}^{\sigma_{-}}=-\delta f_{368}^{\sigma_{+}}$. Ultimately, the observed frequency splitting is conveniently `amplified' more than fourfold without loss of accuracy via this two-step technique and can be expressed in terms of the Land\'{e} $g$-factor, $g_{J}$, the Bohr magneton, $\mu_{\mathrm{B}}$, the applied field, $B$, and Planck's constant, $h$, as
\begin{equation} 
\Delta f=(2) \delta f_{z}\left( 1+\frac{\lambda_{1279}}{\lambda_{368}}\right)=\frac{2g_{J}\mu_{B}B}{h}\left(1+\frac{\lambda_{1279}}{\lambda_{368}} \right).  
\notag 
\end{equation}
For the $(6p^{2})^{3}P_{1}$ level $g_{J}$ = 1.501(2), and the magnetic field can thus be inferred from
\begin{equation} \tag{B.2}
B=\frac{h\Delta f}{2 g_{J} \mu_{B} \left(1+\frac{\lambda_{1279}}{\lambda_{368}} \right)}.    
\end{equation}
$B$ was determined to be linearly related to the applied current, with a slope of $22.0(2)$ G/A and fitted intercept consistent with zero. A room temperature measurement made with a magnetometer probe was found to be consistent with the {\it{in situ}} measurement.

\appendix*
\section{C: Line shape fitting functions}
The line shapes for absorption (Eqs. 1 and 2 and Faraday rotation (Eq. 7) have no analytical solution, and so approximative methods are used to fit our spectral data. For this work, we represent the relevant convolutions as an infinite series of $\gamma=\Gamma/2\sqrt{2}\sigma$, where $\Gamma$ and $\sigma$ are the Lorentzian (FWHM) and Gaussian ($1/e$) spectral widths. The coefficients of these series expansions are expressed in terms of Dawson functions, $F(x)$, \cite{MeekhofThesis} which are built into the MATLAB mathematical toolbox. Here, $F(x)$ is the solution to Dawson's equation, $F'(x)+2xF(x)=1$, where $x$ is the reduced frequency $x=\omega/\sqrt{2}\sigma$. The transmission spectrum is given by
\begin{equation}\tag{C.1}
I(\omega)=I_{0}\exp \left( -\alpha \sum_{k=0}^{\infty}a_{k}\gamma^{k}\right),
\end{equation}
where the leading terms are $a_{0}=\sqrt{\pi}e^{-x^{2}}$, $a_{1}=4xF(x)-2$, $a_{2}=\sqrt{\pi}e^{-x^{2}}(1-2x^{2})$, $a_{3}=-\frac{1}{6}(8-24xF(x)-8x^{2}+16x^{3}F(x))$,...
\\
\indent The Faraday rotation spectrum $-$ which depends on the real part of the refractive index of the atomic vapor $-$ has a different series expansion. Since the optical rotation is proportional to the difference between dispersion curves, the appropriate fitting function is
\begin{equation} \tag{C.2}
\Phi_{F}(\omega)=\frac{C}{\sqrt{2\pi}\sigma}\left(D(\omega+\delta \omega_{z})- D(\omega-\delta \omega_{z})\right),   
\end{equation}
where $\delta \omega_{z}$ is the Zeeman shift, and 
\begin{equation} \tag{C.3}
D(\omega)=\sum_{k=0}^{\infty}d_{k}\gamma^{k}.    
\end{equation}
The coefficients here are given by $d_{0}=2F(x)$, $d_{1}=-2\sqrt{\pi}x e^{-x^{2}}$, $d_{2}=-\frac{1}{2}\left(4F(x)-8x^{2}F(x)+4x \right)$, $d_{3}=\frac{2\sqrt{\pi}}{3}e^{-x^{2}}\left(-3x+2x^{3} \right)$, ... For our line shapes, $\gamma \approx 0.1$ in which case the Dawson expansion converges rapidly and we find that the 3-term expansion differs from an exact numerical convolution by less than 0.01\%.

\end{document}